# Artificial Intelligence for EU Decision-Making: Effects on Citizens' Perceptions of Input, Throughput & Output Legitimacy

Christopher Starke [1]* and Marco Lünich[2]

[1] University of Düsseldorf, Düsseldorf, Germany

[2] University of Düsseldorf, Düsseldorf, Germany

*Corresponding author: christopher.starke@uni-duesseldorf.de

Abstract

A lack of political legitimacy undermines the ability of the European Union (EU) to resolve major crises and threatens the stability of the system as a whole. By integrating digital data into political processes, the EU seeks to base decision-making increasingly on sound empirical evidence. In particular, artificial intelligence (AI) systems have the potential to increase political legitimacy by identifying pressing societal issues, forecasting potential policy outcomes, informing the policy process, and evaluating policy effectiveness. This paper investigates how citizens' perceptions of EU input, throughput, and output legitimacy are influenced by three distinct decision-making arrangements: (1) independent human decision-making (HDM); (2) independent algorithmic decision-making (ADM) by AI-based systems; and (3) hybrid decision-making by EU politicians and AI-based systems together. The results of a pre-registered online experiment (n = 572) suggest that existing EU decision-making arrangements are still perceived as the most democratic (input legitimacy). However, regarding the decision-making process itself (throughput legitimacy) and its policy outcomes (output legitimacy), no difference was observed between the status quo and hybrid decision-making involving both ADM and democratically elected EU institutions. Where ADM systems are the sole decision-maker, respondents tend to perceive these as illegitimate. The paper discusses the implications of these findings for (a) EU legitimacy and (b) data-driven policy-making.

Policy Significance

The results of this experimental study suggest that respondents perceive demanding forms of algorithmic decision-making (ADM) to be illegitimate for EU policy-making. EU policy-makers should exercise caution when incorporating ADM systems in the political decision-making process. ADM systems for far reaching decisions such as budgeting should only be used to assist or consult human decision-makers rather than replacing them. An additional takeaway from this study is that the factual and perceived legitimacy of ADM do not necessarily correspond—that is, even ADM systems that produce high quality outputs and are implemented transparently and fairly may still be perceived as illegitimate and might therefore be rejected by the electorate. To be socially acceptable, implementation of ADM systems must therefore take account of both factual and perceived legitimacy.

## 1. Introduction

The European Union (EU) currently faces a number of significant crises, most notably the European debt crisis, the distribution of refugees across EU member states, and the so-called "Brexit" (withdrawal of the United Kingdom from the EU). As a result, right-wing populist parties promoting anti-EU messages have gained momentum and threaten the stability of the EU as a whole (Schmidt, 2015). To resolve these crises, the EU must demonstrate responsiveness to citizens' concerns (*input legitimacy*), effective and transparent procedures (*throughput legitimacy*) and good governance performance (*output legitimacy*) (Schmidt, 2013; Weiler, 2012). However, the EU allegedly lacks legitimacy on all three counts because of a democratic deficit in the institution's design, the lack of a European identity, and the inadequacies of the European public sphere (Follesdal, 2006; Habermas, 2009; Risse, 2014).

To improve their legitimacy, EU political institutions have increasingly committed to data-driven forms of governance. By integrating digital data into political processes, the EU seeks increasingly to base decision-making on sound empirical evidence (e.g., the Data4Policy Project). In particular, algorithmic decision-making (ADM) systems are used to identify pressing societal issues, to forecast potential policy outcomes, to inform the policy process, and to evaluate policy effectiveness (AlgorithmWatch, 2019; Poel, Meyer, & Schroeder, 2018). For instance, ADM systems have been shown to successfully support decision-making regarding the socially acceptable distribution of refugees. Trials suggest that this approach increases refugee employment rates by 40–70% as compared to human-led distribution practices (Bansak et al., 2018).

However, little is known about the specific impact of ADM on public perceptions of legitimacy (De Fine Licht & De Fine Licht, 2020). On the one hand, high public support for digitalization in general and autonomous systems in particular means that the use of ADM may increase perceived legitimacy (European Commission, 2017). Notably, ADM systems are commonly perceived as true, objective, and accurate and therefore capable of reducing human bias in the decision-making process (Barocas & Selbst, 2016; Lee, 2018). On the other hand, ADM-based policy-making poses a number of novel challenges in terms of perceived legitimacy. (1) Citizens may believe that they have little influence on ADM selection criteria—for instance, which digital data are collected, or on which indicators the algorithm ultimately bases decisions (*input legitimacy*). (2) Citizens may not understand the complex and often opaque technicalities of the ADM process (*throughput legitimacy*). (3) Citizens may doubt that ADM systems can make better decisions than humans, or they may question whether certain decisions produce the desired results (*output legitimacy*).

Few studies have investigated the effects of ADM on perceptions of legitimacy, especially with respect to political decisions. To date, empirical studies have tended to focus on public sector areas such as education and health, evaluating the effects of ADM as compared to human decision-making (HDM) in terms of variables such as fairness and trust (Araujo, Helberger, Kruikemeier, & de Vreese, 2020; Lee, 2018; Marcinkowski, Kieslich, Starke, & Lünich, 2020). To bridge this research gap, the present study investigates the extent to which the use of ADM influences the perceived legitimacy of policy-making at EU level. In so doing, the study extends the existing literature in three respects. (1) It provides novel insights into the potential of ADM to exacerbate or alleviate the EU's perceived legitimacy deficit. (2) It clarifies the effects of three distinct decision-making arrangements on perceptions of legitimacy: (a) independent decision making by EU politicians or "human decision-making" (HDM); (b) independent decision making by ADM systems or "algorithmic decision-making" (ADM); and (c) hybrid decision-making, where politicians select among decisions suggested by ADM systems. (3) Using structural means modeling (SMM) to analyze citizens' perceptions, the study proposes a general measure of input, throughput, and output legitimacy.

## 2. A Crisis of EU Legitimacy? Input, Throughput, Output

In making effective decisions to resolve major crises, the EU's actions depend on political legitimacy. According to Gurr, "governance can be considered legitimate in so far as its subjects regard it as proper and deserving of support" (1971, p. 185). In his seminal work on legitimacy, Scharpf (1999) distinguished between two dimensions of legitimacy; *input legitimacy,* which Schmidt characterized as "responsiveness to citizen concerns as a result of participation *by* the people" (2013, p. 2). Input legitimacy depends on free and fair elections, high voter turnout, and lively political debate in the public sphere (Scharpf, 1999). *Output legitimacy* refers to "the effectiveness of the EU's policy outcomes *for* the people" (Schmidt, 2013, p. 2)—that is, the EU's problem-solving capacity in pursuing desired goals such as preserving peace, ensuring security, protecting the environment, and fostering prosperity (Follesdal, 2006). Moving beyond this dichotomy, some scholars (Schmidt, 2013; Schmidt & Wood, 2019) have added throughput as a third dimension of legitimacy, referring to the accountability, efficacy, and transparency of EU policy-makers and their "inclusiveness and openness to consultation *with* the people" (Schmidt, 2013, p. 2). Also referred to as the "black box" (Steffek, 2018, p. 1), throughput legitimacy encompasses the political practices and processes of EU institutions in turning citizen input into policy output (Schmidt & Wood, 2019; Steffek, 2018).

Ever since the EU was founded, and especially since the failed Constitutional Treaty referenda in France and the Netherlands in 2005, European integration has been dogged by criticisms that the EU lacks legitimacy. Most scholars point to the democratic deficit, the lack of a European identity, and an inadequate public sphere as primary reasons for this alleged crisis of legitimacy (Follesdal & Hix, 2006; Habermas, 2009). The debate centers on four arguments (De Angelis, 2017; Follesdal, 2006; Follesdal & Hix, 2006; Holzhacker, 2007). First, among key EU political institutions, only the European Parliament (EP) is legitimized by European citizens by means of elections, but scholars argue that the EP is too weak in comparison to the European Commission (EC) (Follesdal & Hix, 2006). While continuous reform of EU treaties has substantially strengthened the EP's role within the institutional design of the EU, it still lacks the power to initiate legislation (Holzhacker, 2007). Second, the EU's institutional design gives national governments pivotal power over the Council of the EU and the EC. However, as those actors are somewhat exempt from parliamentary scrutiny by the EP and national parliaments, there is a deficit in democratic checks and balances (Follesdal & Hix, 2006). Third, the European elections are not sufficiently "European" (Follesdal, 2006)—that is, "they are not about the personalities and parties at the European level or the direction of the EU policy agenda" (Follesdal & Hix, 2006, p. 536). Instead, national politicians, parties, and issues still dominate campaigns and remain crucial in citizens' voting decisions (Hobolt & Wittrock, 2011). Finally, "the EU is simply 'too distant' from voters" (Follesdal & Hix, 2006, p. 536). Public opinion research suggests that although a sense of European identity, trust in European institutions, and satisfaction with EU democracy are on the rise, these pale in comparison to the corresponding scores at national level (European Commission, 2019b; Risse, 2014). Consequently, scholars have argued that the EU lacks a European demos—that is, "a strong sense of community and loyalty among a political group" (Risse, 2014, p. 1207). In addition, the alleged lack of a European public sphere that would enable communication and debate around political issues lends further credence to the claim that the EU suffers from insufficient citizen participation (Habermas, 2009; Kleinen-von Königslöw, 2012).

As all four arguments primarily question the EU's input and throughput legitimacy, many have argued that output is the stronghold for the EU legitimacy. According to Scharpf, "the EU has developed considerable effectiveness as a regulatory authority" (2009, p. 177). In that regard, the EU enables member states to implement policies that they would otherwise be unable to advance, especially in relation to global policy issues (Menon & Weatherill, 2008). Weiler contended that output legitimacy "is part of the very ethos of the Commission" (2012, p. 828), but recent crises have also challenged this view; for instance, the austerity measures imposed on debtor states had detrimental effects on the lives of many European citizens (De Angelis, 2017). Debate about the EU's alleged legitimacy crisis centers primarily on institutional shortcomings in the political system, and public perceptions of legitimacy are neglected. However, Jones (2009) claimed that subjective perceptions are often more important than the normative criteria themselves.

## 3. ADM for policy-making in the EU?

In recent years, EU institutions have increasingly sought to address this perceived deficit of legitimacy through evidence-based policy-making: "Against the backdrop of multiple crises, policymakers seem ever more inclined to legitimize specific ways of action by referring to 'hard' scientific evidence suggesting that a particular initiative will eventually yield the desired outcomes" (Rieder & Simon, 2016, p. 1). This push for numerical evidence comes at a time when the computerization of society has precipitated the creation and storage of vast amounts of digital data. According to boyd and Crawford, so-called big data "offer a higher form of intelligence and knowledge that can generate insights that were previously impossible, with the aura of truth, objectivity, and accuracy" (2012, p. 663). The value assigned to big data as an information asset reflects the insights gained from characteristics such as volume, variety, velocity, and presumed veracity (De Mauro, Greco, & Grimaldi, 2016). Digital data are collected, accessed, and analyzed in real time, leading to substantial advances in analytics, modeling, and dynamic visualization (Craglia et al., 2018; Poel et al., 2018). This transformation of real-world phenomena into digital data is expected to provide a timely and undistorted view of societal mechanisms and institutions.

Lately, public discourse around the potential of computerization and big data has included a renewed focus on Artificial Intelligence (AI). According to Katz,

> "AI stands for a confused mix of terms—such as 'big data,' 'machine learning,' or 'deep learning'—whose common denominator is the use of expensive computing power to analyze massive centralized data. (…) It's a vision in which truth emerges from big data, where more metrics always need to be imposed upon human endeavors, and where inexorable progress in technology can 'solve' humanity's problems" (2017, p. 2).

Indeed, the increasing availability of digital data in combination with significant advances in computing power have underpinned the recent emergence of many successful AI applications such as self-driving cars and automated text production and face recognition. This has in turn raised expectations regarding the use of AI for evidence-based or data-driven policy-making (Esty & Rushing, 2007; Giest, 2017; Poel et al., 2018). To exploit technological developments and increasing data availability for policy-making purposes, the EC introduced the Data4Policy initiative (European Commission, 2019a), arguing that "data technologies are amongst the valuable tools that policymakers have at hand for informing the policy process, from identifying issues, to designing their intervention and monitoring results" (European Commission, 2019a, para. 1).

In that context, van Veenstra and Kotterink (2017, p. 101) noted that "data-driven policy making is not only expected to result in better policies, but also aims to create legitimacy." Recent reports suggest that "algorithmically driven, automated decision-making (ADM) systems are already in use all over the EU" (AlgorithmWatch, 2019, p. 10) to deliver public services, optimize traffic flows, or identify social fraud (AlgorithmWatch, 2019; Poel et al., 2018). Case studies confirm that ADM systems can indeed contribute to better policy (Bansak et al., 2018), using big data to identify emerging issues, to foresee demand for political action, to monitor social problems, and to design policy options (Poel et al., 2018; Verhulst, Engin, & Crowcroft, 2019). To that extent, data-driven systems can potentially contribute to the increased legitimacy of *input* (by enabling new forms of citizen participation,) of *throughput* (by making the political process more transparent), and of *output* (by increasing the quality of policies and outcomes).

Yet, despite these promising indications, there are numerous examples of AI's downside in political decision-making. For instance, a recent report by the research institute AI NOW revealed that ADM systems may falsely accuse citizens of fraud, arbitrarily exclude them from food support programs, or mistakenly reduce their disability benefits. Incorrect classification by ADM systems has led to a wave of lawsuits against the US government at federal and state levels, undermining both the much vaunted cost efficiency of automated systems and the perceived legitimacy of political decision-making as a whole (Richardson, Schultz, & Southerland, 2019). In terms of the three dimensions of legitimacy, ADM systems pose the following challenges. (1) On the input dimension, citizens may lack insight into or influence over the criteria or data that intelligent algorithms use to make decisions. (2) On the throughput dimension, citizens may be unable to comprehend the complex and often inscrutable logic that underpins algorithmic predictions, recommendations, or decisions. (3) On the output dimension, citizens may fundamentally doubt whether ADM systems actually contribute to more efficient policy.

As with all technological innovations, success or failure depends greatly on all stakeholders' participation and acceptance (Bauer, 1995). In the present context, those stakeholders include EU institutions and bureaucracies, operatives and regulators who may favor the introduction of ADM systems in policy-making, and the electoral body of voters and citizens who lend legitimacy to proposed policies and implementation. The present study focuses on perceptions of European citizens regarding the legitimacy of using ADM systems in EU governance and political decision-making.

While there are no existing accounts of citizens' perceptions of ADM systems in the context of political decision-making, survey data provide some initial insights. Several Eurobarometer surveys have shown that public perception of digital technologies is broadly positive throughout the EU, especially when compared to perceptions of other mega-technologies such as nuclear power, biotechnology, or gene editing (European Commission, 2015, 2017). According to a recent survey commissioned by the Center for the Governance of Change, "25% of Europeans are somewhat or totally in favor of letting an artificial intelligence make important decisions about the running of their country" (Rubio & Lastra, 2019, p. 10). On that basis, it seems likely that demands to embed AI in the political process will increase, and that political programs will respond to those demands.

## 4. Hypotheses

The key objective of this study was to investigate whether and to what extent ADM systems in policy-making influence public perceptions of EU input, throughput, and output legitimacy. More specifically, we examined the decision-making process that determines how the EU's annual budget is distributed. Previous empirical studies have suggested that different decision-making arrangements (e.g., formal vs. descriptive representation, direct voting vs. deliberation) can differ significantly in terms of their perceived legitimacy (Arnesen, Broderstad, Johannesson, & Linde, 2019; Arnesen & Peters, 2018; Esaiasson, Gilljam, & Persson, 2012; Persson, Esaiasson, & Gilljam, 2013). However, as those studies did not specifically investigate the potential effects of ADM systems, the present study sought to distinguish between three different decision-making arrangements: (1) independent decision-making by EU politicians (*human decision-making*); (2) independent decision-making by ADM systems (*algorithmic decision-making*); and (3) *hybrid decision-making* by politicians, based on suggestions made by ADM systems.

With regard to perceived input legitimacy, we contend that respondents are likely to perceive the current decision-making process as more legitimate than processes that rely partly or completely on ADM. The primary reason for this assumption is that ADM would arguably render citizen participation in EU governance largely obsolete or at least diminished. First attempts suggest that data science could be used to asses public leaning towards policy issues via text mining of Social Media data (Sluban & Battiston, 2017). Yet, using technological solutions to make policy decisions based on such automated monitoring of public opinion is unlikely to increase public perceptions of input legitimacy, especially not at the expense of existing democratic procedures. Thus, even though current policy-making procedures in the EU are often criticized for their democratic deficit, for the foreseeable future algorithmic or hybrid decision systems would likely marginalize citizen participation, especially if they are used for decision-making rather than less demanding forms of the policy cycle such as agenda setting or evaluation (Verhulst et al., 2019). To that extent, the introduction of ADM would weaken the perceived democratic quality of political action in the EU. On that basis, we tested the following pre-registered hypotheses (see [pre-registration](#) at OSF):

H1a: Human decision-making leads to higher perceived input legitimacy as compared to algorithmic decision-making.

H1b: Human decision-making leads to higher perceived input legitimacy as compared to hybrid decision-making.

H1c: Hybrid decision-making leads to higher perceived input legitimacy as compared to algorithmic decision-making.

With regard to perceived throughput legitimacy, we argue that implementation of ADM leads to lower levels of perceived legitimacy as compared to the existing political process. While EU decision-making processes are often criticized for their lack of transparency, ADM systems suffer from the same deficiency, as they are themselves considered to be a "black box" (Wachter, Mittelstadt, & Russell, 2018). The extent of transparency of self-learning systems, however, is a major driver of public perceptions of legitimacy ((De Fine Licht & De Fine Licht, 2020). A recent EC report therefore stressed the urgent need to make ADM more explainable and transparent (Craglia et al., 2018), on the grounds that such systems are typically too complex for the layperson to understand and are largely unable to give proper justifications for decisions. They further lack public accountability because citizens do not know who to turn to regarding policy or administrative failures. Indeed, preliminary empirical evidence suggests that activities that require human skills are perceived as fairer and more trustworthy when executed by humans rather than algorithms (Lee, 2018). On that basis, we formulated the following hypotheses.

H2a: Human decision-making leads to higher perceived throughput legitimacy as compared to algorithmic decision-making.

H2b: Human decision-making leads to higher perceived throughput legitimacy as compared to hybrid decision-making.

H2c: Hybrid decision-making leads to higher perceived throughput legitimacy as compared to algorithmic decision-making.

Several scholars suggest that the EU already legitimizes itself primarily via the output dimension (Scharpf, 2009; Weiler, 2012) due to the aforementioned democratic decifit on the input dimension. Below, we argue why implementing algorithmic or hybrid decision system would mean that the EU is doubling down on ouput legitimacy. Perceived output legitimacy comprises two key dimensions: citizens' perceptions of whether political decisions can attain predefined goals (e.g., economic growth, environmental sustainability), and the subjective favorability of such decisions. Assessment of the perceived quality of political output involves both dimensions, and this is where ADM systems are said to have a distinct advantage over human decision-makers, as they can produce novel insights from vast amounts of digital data that would be impossible when relying solely on human intelligence (boyd & Crawford, 2012). Empirical studies comparing public perceptions of ADM and HDM seem to support this assumption; looking at proxies for legitimacy, ADM systems are evaluated as fairer in distributive terms than HDM (Marcinkowski et al., 2020), especially in high impact situations (Araujo et al., 2020). Building on these empirical findings, we further argue that citizens perceive ADM systems to be most legitimate when they operate under the scrutiny of democratically elected institutions. Thus, we formulated the following hypotheses.

H3a: Human decision-making leads to lower perceived goal attainment as compared to algorithmic decision-making.

H3b: Human decision-making leads to lower perceived goal attainment as compared to hybrid decision-making.

H3c: Hybrid decision-making leads to higher perceived goal attainment as compared to algorithmic decision-making.

H4a: Human decision-making leads to lower decision favorability as compared to algorithmic decision-making.

H4b: Human decision-making leads to lower decision favorability as compared to hybrid decision-making.

H4c: Hybrid decision-making leads to higher decision favorability as compared to algorithmic decision-making.

## 5. Method

To test these hypotheses, we conducted an online experiment, applying a between-subjects-design using one factor with three levels: (1) EU politicians making decisions independently (cond$^{HDM}$); (2) ADM systems making decisions independently (cond$^{ADM}$); and (3) ADM systems suggesting decisions to be passed by EU politicians (cond$^{Hybrid}$) (see pre-registration[1] at OSF). All measurements and stimulus material and the questionnaire's basic functionality were thoroughly tested in multiple pre-tests involving 321 respondents in total.

### 5.1. Sample

Respondents were recruited through the non-commercial SoSci Open Access Panel (OAP) during the period 8–22 April 2019. In accordance with German law, SoSci OAP registration involves a double opt-in process, in which panelists first sign up using an email address and must then activate their account and confirm pool membership (Leiner, 2016). Although the SoSci OAP is not representative in terms of socio-demographic variables, its key advantage is participant motivation; as respondents are not compensated for survey participation, their main motivation is topic interest, which is a crucial indicator of data quality (Brüggen, Wetzels, De Ruyter, & Schillewaert, 2011). In addition, all questionnaires using the SoSci OAP must first undergo rigorous peer review, so ensuring "major improvements to the instrument before data is collected" (Leiner, 2016, p. 373).

Using Soper's (2019) a priori sample size calculator for structural equation modeling, we determined an optimal sample size of 520, based on the results from a pre-test conducted 10 weeks before final data collection. In total, 612 respondents completed the questionnaire. A thorough two-step cleaning process was applied for quality control purposes. The first step excluded respondents who failed an attention check regarding the target topic ($n = 14$). In the second step, using the DEG_Time variable (Leiner, 2013), each respondent accumulated minus points for completing single questions or the whole questionnaire too quickly. As the SoSci OAP administrators recommend a threshold score of 50 for rigorous filtering, all respondents with a minus point score of 50 or higher were excluded from the analysis ($n = 26$). After filtering, the sample comprised 572 respondents—a response rate of 19.1%. No differences were observed between the conditions in terms of age ($M = 47.26$, $SD = 15.99$; $F(2, 569) = .182$, $p = .834$); gender (female = 45.8 %, male = 53.5 %, diverse = .7 %; $X^2(4) = 1.89$, $p = .757$); and education (non-tertiary education = 40.6 %, tertiary education = 59.4 %; $X^2(2) = .844$, $p = .656$).

---

[1] Table 4 in the Appendix accounts for all deviations from pre-registration.

### 5.2. Treatment conditions (independent variable)

Respondents were randomly assigned to one of the three conditions and received a short text (ca. 250 words per condition) about the decision-making process regarding distribution of the annual EU budget. The stimulus material also included a pie chart showing budget allocation for different policy areas. The text was adapted from the official EU website (European Union, 2019). While the pie chart was identical for all three conditions, the closing paragraph of the text was edited to reflect manipulation of the independent variable: (a) decisions made by politicians of EU institutions only—the status quo ($cond^{HDM}$; $n = 182$); (b) decisions made by ADM only ($cond^{ADM}$; $n = 204$); and (c) decisions suggested by ADM and subsequently passed by politicians of EU institutions ($cond^{Hybrid}$; $n = 186$)[2]. Respondents were <u>not</u> deceived into thinking that $cond^{ADM}$ and $cond^{Hybrid}$ are existing decision-making procedures in the EU as it was explicitly stressed that the scenario at hand was only a *potential* decision-making process. At the end of the survey, they were debriefed about the research interest of the study

### 5.3. Manipulation Check

All respondents answered two items that served as manipulation checks to validate that respondents perceived the differences in the respective conditions. First, perceived technical automation of the decision-making process was assessed by responses (on a 5-point Likert scale) to the question *How technically automated was the decision-making process?* The results indicated a significant difference among the three conditions ($F(2, 524) = 389.71$ $p < .001$). Using a Games-Howell post hoc test, $cond^{HDM}$ ($M = 2.11$; $SD = .97$), $cond^{ADM}$ ($M = 4.52$; $SD = .77$), and $cond^{Hybrid}$ ($M = 4.16$; $SD = .80$) were found to differ significantly from each other, confirming that respondents recognized the extent to which the described decision-making processes were technically automated.

The perceived involvement of political actors and institutions in the different decision-making arrangements was measured by responses (on a 5-point Likert scale) to the question *What role did politicians or political institutions play in the decision-making process?* Again, there were significant differences among the three conditions ($F(2, 548) = 161.98$, $p < .001$). Using a Games-Howell post hoc test, $cond^{HDM}$ ($M = 4.45$; $SD = .88$), $cond^{ADM}$ ($M = 2.63$; $SD = .99$), and $cond^{Hybrid}$ ($M = 3.41$; $SD = 1.04$), all were found to differ significantly from each other, confirming that respondents recognized the degree to which political actors and institutions were involved in each condition.

### 5.4. Measures

As Persson et al. (2013, p. 391) rightly noted, "legitimacy is an inherently abstract concept that is hard to measure directly." To account for this difficulty, measures for *input legitimacy* (dV1), *throughput legitimacy* (dV2) and output legitimacy using the two dependent variables *goal attainment* (dV3) and *decision favorability* (dV4) were thoroughly pre-tested and validated. All items used in the analysis were measured on a 5-point Likert scale, ranging from 1 (*do not agree*) to 5 (*agree*) and including the residual category *don't know*. The factor validity of all measures was assessed using Cronbach's alpha (α) and average variance extracted (AVE). Means (*M*), standard deviations (*SD*), α scores, and AVE of all measures are shown in Table 1.

---

[2] A translation of the stimulus material can be found in the Appendix.

**Table 1.**
*Descriptives and Factorial Validity*

|                       | M    | SD   |
|-----------------------|------|------|
| Input Legitimacy      | 1.66 | 0.85 |
| Throughput Legitimacy | 2.76 | 1.00 |
| Goal Attainment       | 3.28 | 0.91 |
| Decision Acceptance   | 2.96 | 1.05 |

*Note.* AVE = Average Variance Extracted

*Input Legitimacy* (dV1). Three items were used to measure perceived input legitimacy, using wording adapted[3] from previous studies (Colquitt & Rodell, 2015; Lindgren & Persson, 2010; Persson et al., 2013): (1) *All citizens had the opportunity to participate in the decision-making process* (IL1); (2) *People like me could voice their opinions in the decision-making process* (IL2)*; and* (3) *People like me could influence the decision-making process* (IL3). All items were randomized and used as indicators of a latent variable in the analysis.

*Throughput Legitimacy* (dV2). To measure perceived throughput legitimacy, three items were adapted from Werner and Marien (2018). Respondents were asked indicate to what extent they perceived the decision-making process described in the stimulus material as (1) *fair* (TL1); (2) *satisfactory* (TL2); and (3) *appropriate* (TL3). All items were randomized and used as indicators of a latent variable in the analysis.

*Goal Attainment* (dV3). To measure perceived goal attainment, which is considered an important pillar of output legitimacy (Lindgren & Persson, 2010), respondents were asked to indicate to what extent they believed the decision-making process could achieve the goals referred to in the stimulus text (adapted from the official EU website): (1) *Better development of transport routes, energy networks and communication links between EU countries* (GA1);. (2) *Improved protection of the environment throughout Europe* (GA2); (3) *An increase in the global competitiveness of the European economy* (GA3); and (4) *Promoting cross-border associations of European scientists and researchers* (GA4) (European Union, 2019). The order of the items was randomized. As the four goals can be independently attained, the underlying construct is not one-dimensional and reflective. For that reason, we computed a mean index for goal attainment that was used as a manifest variable in the analysis.

*Decision Favorability* (dV4). In the existing literature, decision acceptance or favorability is commonly used as a measure of legitimacy (Esaiasson et al., 2012; Werner & Marien, 2018). Conceptualizing decision favorability as the second key pillar of output legitimacy, we used three items to measure dV4. Two of these items were adopted from Werner and Marien's (2018) four-item scale: (1) *I accept the decision* (DF1), and (2) *I agree with the decision* (DF2). As the other two items in their scale refer to the concept of reactance, we opted to formulate one additional item: (3) *The decision satisfies me* (DF3). All items were randomized and used as indicators of a latent variable in the analysis.

### 5.5. Data Analysis

The analysis employed structured means modeling (SMM), incorporating all four variables in a single model. As this approach takes account of measurement error due to latent variables, it was adopted in preference to traditional analysis of variance (Breitsohl, 2018). To test the hypotheses, we compared means between groups, using critical ratios for differences between parameters in the specified model. All statistical analyses were performed using AMOS 23. Because of missing data, Full Information Maximum Likelihood estimation was used in conjunction with estimation of means and intercepts (Kline, 2016). Full model fit was assessed using a chi-square test and RMSEA (lower and upper bound of the 90% confidence interval, PClose value), along with the Tucker-Lewis-Index (TLI) measure of goodness of fit (Holbert & Stephenson, 2002; van de Schoot, Lugtig, & Hox, 2012). Differences in means were investigated by obtaining critical ratios (CR); for CR > 1.96 or < -1.96, respectively, the parameter difference indicated two-sided statistical significance at the 5% level.

---

[3] All items were translated from English into German.

As the experimental design compared three groups, we tested the measurement models of all latent factors for measurement invariance (Kline, 2016; van de Schoot et al., 2012). This test was necessary to assess whether factor loadings (metric invariance) and item intercepts (scalar invariance) were equal across groups. This "strong invariance" is a necessary precondition to confirm that latent factors are measuring the same construct and can be meaningfully compared across groups (Widaman & Reise, 1997). The chi-square-difference test for strong measurement invariance in Table 2 shows that the assumptions of metric and scalar invariance are violated. Subsequent testing of indicator items identified indicator IL02 as non-invariant. On that basis, a model with only partial invariance was estimated, freeing both the indicator loading and item intercept constraints of IL01. A chi-square-difference test for partial measurement invariance showed better model fit as compared to the configural model ($\Delta X^2$ = 18.034, $\Delta df$ = 16; $p$ = .322). The final model with partial measurement invariance fit the data well ($X^2$(106) = 173.299, $p < .001$; RMSEA = .033 (.024; .042); PClose = .999; TLI = .966). The latent means of the specified model with partial invariance were constrained to zero in $cond^{HDM}$. On that basis, the first condition, in which only EU politicians made decisions about the EU budget, was used as the reference group when reporting the results of group comparisons.

**Table 2.**
*Descriptives and Factorial Validity*

|  | $X^2$ | df | p | TLI | RMSEA | PClose |
|---|---|---|---|---|---|---|
| Configural Model | 155.265 | 90 | < .001 | .961 | .036 (.026; .045) | .995 |
| Metric Invariance | 198.964 | 102 | < .001 | .949 | .041 (.032; .049) | .963 |
| Scalar Invariance | 219.483 | 114 | < .001 | .950 | .040 (.032; .048) | .978 |
| Final Model with Partial Invariance | 173.299 | 106 | < .001 | .966 | .033 (.024; .042) | .999 |

## 6. Results

Construct means are shown in Table 3. In addition, based on a transformation of Hedge's g, a standardized effect size r as proposed by Steimetz et al. (2009) was manually calculated. This is also reported in Table 3.

**Table 3.**
*Comparisons of Structured Means of the Legitimacy Dimensions*

|  | Means | | | Effect Sizes (r) | | |
|---|---|---|---|---|---|---|
|  | $cond^{HDM}$ | $cond^{ADM}$ | $cond^{Hybrid}$ | $cond^{HDM}$ vs. $cond^{ADM}$ | $cond^{ADM}$ vs. $cond^{Hybrid}$ | $cond^{HDM}$ vs. $cond^{Hybrid}$ |
| Input Legitimacy | 0 | -.494 | -.262 | .25 | .13 | .13 |
| Throughput Legitimacy | 0a | -.346 | .070a | .15 | .19 | .00 |
| Goal Attainment | 3.37a | 3.15b | 3.29a,b, | .11 | .00 | .00 |
| Decision Acceptance | 0a | -.347 | .025a | .15 | .15 | .00 |

*Note.* Means not sharing any letter are significantly different by the test of critical ratios at the 5% level of significance.

With regard to perceived input legitimacy, we assumed that this would be highest in $cond^{HDM}$ (in which only EU politicians made budget decisions) and lowest in $cond^{ADM}$ (decisions based solely on ADM), with $cond^{Hybrid}$ (ADM and EU politicians combined) somewhere between the two. The results indicate that respondents perceived input legitimacy as significantly lower in $cond^{ADM}$ ($\Delta M$ = -.494, $p < .001$) and $cond^{Hybrid}$ ($\Delta M$ = -.262, $p$ = .011). As the difference between these conditions was also significant ($\Delta M$ = -.232, $p$ = .009), hypotheses H1a, H1b, and H1c were supported.

For perceived throughput legitimacy, the results indicate (as expected) that $cond^{ADM}$ was perceived as significantly less legitimate than $cond^{HDM}$ ($\Delta M$ = -.346, $p < .001$). No difference was observed between $cond^{HDM}$ and $cond^{Hybrid}$ ($\Delta M$ = -.070, $p$ = .481), but $cond^{Hybrid}$ differed significantly from $cond^{ADM}$ ($\Delta M$ = -.276, $p$ = .004). As a consequence, hypotheses H2a and H2b were supported while H2c was rejected.

In contrast to input and throughput legitimacy, we assumed that cond$^{HDM}$ would score lower than the other two conditions for perceived goal attainment, and that cond$^{Hybrid}$ would score higher than the other two conditions. In fact, cond$^{HDM}$ returned the highest mean ($M = 3.37$) and did not differ significantly from cond$^{Hybrid}$ ($M = 3.29$; $\Delta M = .083$, $p = .383$). Again, cond$^{ADM}$ scored lowest ($M = 3.15$) and differed significantly from cond$^{HDM}$ ($\Delta M = .223$, $p = .014$) but not from cond$^{Hybrid}$ ($\Delta M = -.014$, $p = .151$). These results found no support for hypotheses H3a, H3b, or H3c and even ran counter to the assumptions of H3a.

We anticipated that perceived decision favorability would be highest for cond$^{Hybrid}$, lowest for cond$^{HDM}$, with cond$^{ADM}$ somewhere between the two. In fact, cond$^{ADM}$ scored significantly lower than cond$^{HDM}$ ($\Delta M = -.347$, $p < .001$) and significantly lower than cond$^{Hybrid}$ ($\Delta M = -.372$, $p < .001$). There was no significant difference between cond$^{HDM}$ and cond$^{Hybrid}$ ($\Delta M = -.25$, $p = .809$). As a result, H4a and H4b were rejected while H4c was accepted.

## 7. Discussion

This paper answers the call for more empirical research to understand the nexus of ADM for political decision-making and its perceived legitimacy (De Fine Licht & De Fine Licht, 2020). How does the integration of AI into political decision-making influence people's perceptions of the legitimacy of the decision-making process? In pursuit of preliminary answers to this question, the results of a pre-registered online experiment that systematically manipulated levels of algorithmic decision-making (ADM) in EU policy-making yielded three main insights. First, existing EU decision-making arrangements were considered the most democratic—that is, they scored highest on input legitimacy. Second, in terms of process quality (throughput legitimacy) and outcome quality (output legitimacy), no differences were observed between existing decision-making arrangements and hybrid decision-making. Finally, decision-making informed solely by ADM was perceived as the least legitimate arrangement across all three dimensions. In the following sections, we consider the implications of these findings for EU legitimacy, data-driven policy-making, and avenues for future research.

### 7.1. Implications for the legitimacy of the EU

Our findings lend further credence to previous assertions that the EU lacks political legitimacy (Holzhacker, 2007), in that current decision-making arrangements, which solely involve EU politicians, score low on input legitimacy ($M = 1.90$ on a 5-point Likert scale). This finding speaks to a previously noted democratic deficit (Follesdal, 2006; Follesdal & Hix, 2006). The present results further reveal that ADM systems do not seem to offer an appropriate remedy; on the contrary, it seems that such systems may even exacerbate the problem, as the existing process is still perceived as having greater input legitimacy than arrangements based wholly or partly on ADM systems. It appears that ADM systems fail to engage citizens in the decision-making process or to make their voices heard. Implementing ADM technologies to assist or replace human political actors is seen as less democratic than the status quo, even though incumbent decision-makers such as the European Commission themselves lack democratic legitimacy. One plausible explanation for this finding is that ADM systems are even more technocratic and detached from voters than EU politicians. For that reason, citizens favor human decision makers when dealing with human tasks, aligning with earlier findings by Lee (2018).

As the EU depends heavily on public approval, it seems important to explore alternative ways of increasing its legitimacy. Rather than leaving political decisions to ADM systems, less demanding forms of data-driven policy-making might help to achieve this goal. Beyond decision making, data-driven applications can help to address input legitimacy deficits by contributing to a much wider range of tasks that include foresight, agenda setting, and evaluation (AlgorithmWatch, 2019; Poel et al., 2018). For instance, some existing applications already use public discourse and opinion poll data to predict issues that require political action before these become problematic. Further empirical investigation is needed to assess how such applications might affect legitimacy perceptions. In relation to ADM, our findings highlight citizens' skepticism regarding the potential of digital technologies to increase input legitimacy in terms of democratic participation, and the EU must assess the use of ADM in this light.

With regard to the quality of decision-making processes—that is, throughput legitimacy—we found no difference between existing decision-making arrangements and hybrid regimes involving ADM systems and EU politicians.

However, citizens seem to view decision making based solely on ADM systems as less fair or appropriate than the other two arrangements. Regarding existing EU procedures and practices, critics lament a lack of transparency, efficiency, and accountability (Schmidt & Wood, 2019), but ADM systems exhibit the same deficiency (Shin & Park, 2019). Inside the "black box", ADM systems change and adapt decision-making criteria according to new inputs and elusive feedback loops that defy explanation even among AI experts. Under the umbrella term "explainable AI," a significant strand of the computer science literature seeks to enhance ADM's transparency to users and the general public (Miller, 2019; Mittelstadt, Russell, & Wachter, 2019). For instance, "counterfactual explanations" indicate which ADM criteria would need to be changed to arrive at a different decision (Wachter et al., 2018).

Regarding citizens' perceptions of the effectiveness and favorability of decision-making outcomes (output legitimacy), we found no difference between the existing decision-making process and hybrid regimes incorporating ADM systems and EU politicians. ADM-based systems alone are considered unable to achieve desired goals, and citizens would not approve of the corresponding decisions. It is important to note that decision output was identical for all three experimental conditions, and that only the decision-making process varied. Nevertheless, these result in differing perceptions of output legitimacy, implying that factual legitimacy (as in the actual quality of policies and their outcomes) and perceived legitimacy are not necessarily congruent. In relation to the European debt crisis, Jones (2009) suggested that political institutions must convince the public that they are performing properly, whatever their actual performance. As the interplay between actual and perceived performance also seems important in the case of ADM, we contend that both aspects warrant equal consideration when implementing such systems in policy-making.

Some of the present results run counter to our hypotheses. Given the largely positive attitude to AI in the EU (European Commission, 2015, 2017), and in light of recent empirical evidence (Araujo et al., 2020; Marcinkowski et al., 2020), we expected ADM to score highly on output legitimacy. However, respondents expressed a more favorable view of HDM and hybrid decision-making outcomes, suggesting that they consider it illegitimate to leave important EU political decisions solely to automated systems. ADM systems were considered legitimate as long as humans remained in the loop, indicating that to maintain existing levels of perceived legitimacy, ADM systems should support or consult human policy-makers rather than replacing them. Furthermore, the results suggest that implementing hybrid systems for EU policy-making would mean that the EU doubles down on focusing on output legitimacy instead of input legitimacy as citizens by no means perceive those systems to solve the democratic deficit of the EU, yet they consider them to produce equally good policy outcomes. As a byproduct ADM or hybrid systems arguably contribute to public perceptions of the EU as a technocractic political system. The findings also indicate that increasing factual legitimacy (e.g., by improving the quality of policy outcomes) does not necessarily yield a corresponding increase in perceived legitimacy.

### 7.2. Implications for data-driven policy-making

Our findings also contribute to the current discussion around data-driven or algorithmic policy-making. To begin, ADM systems do not seem to enhance citizens' assessment of decision-making procedures or outcomes; indeed, the use of ADM systems as sole decision-makers diminishes perceived legitimacy. However, when such systems operate under the scrutiny of democratically elected institutions (as in the hybrid condition), they are seen to be as legitimate as the existing policy-making process. This suggests that including humans in the loop is a necessary precondition for implementing ADM (Goldenfein, 2019), and recent reports indicate that this may be the more plausible scenario in the immediate future (AlgorithmWatch, 2019; Poel et al., 2018). This finding has important implications for data-driven policy-making, as it shows that citizens view human-in-the-loop decision making as legitimate arguably because politicians can modify or overrule decisions made by ADM systems (Dietvorst, Simmons, & Massey, 2018).

Of course, our study tests a very demanding form of algorithmic policy-making in which algorithms take important budgeting decisions under conditions of limited (hybrid condition) or no (ADM condition) democratic oversight. Yet, as Verhulst, Engin and Crowcroft point out: "Data have the potential to transform every part of the policy-making life cycle—agenda setting and needs identification; the search for solutions; prototyping and implementation of solutions; enforcement; and evaluation" (Verhulst et al., 2019, p. 1). Public administration has only recently begun to exploit the potential of ADM to produce better outcomes (Wirtz, Weyerer, & Geyer, 2018). For instance, the Netherlands now uses an ADM system to detect welfare fraud, and in Poland, the Ministry of Justice has implemented an ADM system that randomly allocates court cases to judges (AlgorithmWatch, 2019). Given the increased data availability and computing power fueling powerful AI innovations, it is reasonable to assume that we have only scratched the surface of algorithmic policy-making and that more demanding forms of ADM will be implemented in the future. Moreover, first opinion polls suggest that significant shares of citizens (25% in the EU) agree with AI taking over important political decisions about their country (Rubio & Lastra, 2019). The present findings suggest that implementation processes should be designed to facilitate synergies between algorithmic and human decision-making.

### 7.3. Implications for future empirical research

Three main limitations of this study outline avenues for future empirical research. First, our sample was not representative of the German population. As data were collected using the non-commercial SoSci Open Access Panel (OAP), the convenience sample was skewed in terms of education. This may have yielded slightly more positive perceptions of current EU legitimacy (HDM condition) as compared to the German population, as previous evidence suggests that higher levels of education are associated with more positive attitudes to the EU (Boomgaarden, Schuck, Elenbaas, & de Vreese, 2011). To make stronger claims in terms of the generalizability of the results, future research should use representative national samples.

Second, our study was limited to Germany. While German citizens generally hold more positive views of the EU compared to the European average (European Commission, 2019b), they also favor algorithmic decision-making in politics more than the European average (Rubio & Lastra, 2019). Future studies should investigate the relationship between AI-driven decision-making and perceptions of legitimacy in other national contexts and by means of cross-country comparisons. For instance, preliminary opinion polls suggest that Netherlands citizens express much higher support for ADM in policy-making than citizens of Portugal (43% versus 19%, respectively) (Rubio & Lastra, 2019).

Finally, two of the three decision-making arrangements tested here are hypothetical and are unlikely to be implemented in the immediate future—that is, ADM systems are unlikely to be authorized to allocate the EU's annual budget. On that basis, future research should focus on the effects of less abstract data-driven applications on perceived legitimacy at different stages of the policy cycle and should include varying degress of transparency of self-learning systems (De Fine Licht & De Fine Licht, 2020). For instance, citizens may consider it more legitimate to employ AI-based systems to identify existing societal issues requiring political action or to evaluate the success of legislation based on extensive available data, on the condition that such systems will be able to give convincing justifications for their decisions.

### 8. Conclusion

Conclusions are given here. This study sheds light on citizens' perceptions of the legitimacy of using ADM in EU policy-making. Based on these empirical findings, we suggest that EU policy-makers should exercise caution when incorporating ADM systems in the decision-making process. To maintain current levels of perceived legitimacy, ADM systems should only be used to assist or consult human decision-makers rather than replacing them, as excluding humans from the loop seems detrimental to perceived legitimacy. Second, it seems clear that the factual and perceived legitimacy of ADM do not necessarily correspond—that is, even ADM systems that produce high quality outputs and are implemented transparently and fairly may still be perceived as illegitimate and will therefore be rejected. To be socially acceptable, implementation of ADM systems must therefore take account of both factual and perceived legitimacy.

Abbreviations

| | |
|---|---|
| **EU** | European Union |
| **EP** | European Parliament |
| **EC** | European Commission |
| **AI** | Artificial Intelligence |
| **ADM** | Algorithmic Decision-Making |
| **HDM** | Human Decision-Making |
| **SMM** | Structured Means Modeling |
| **OAP** | Open Access Panel |

Appendix

**Table 4.**

*Deviations from the Pre-Registration*

| Pre-Registration | Paper | Reason |
|---|---|---|
| Terminology: "AI Systems" | Terminology: "ADM Systems" (Algorithmic Decision Making) | ADM is the more precise terminology (AlgorithmWatch, 2019) |
| Measurement of Throughput Legitimacy with six items (fair, satisfactory, just, appropriate, reliable, trustworthy) | Measurement of Throughput Legitimacy with three items (fair, satisfactory, appropriate) | Better fit of the measurement model, all latent constructs were measured with three items |
| Full final model with control variables (Acceptance of technology, political interest, perceived plausibility) | Full final model with control variables (Acceptance of technology, political interest, perceived plausibility) | Better power of the study, better fit indices of the model, control variables were not important for hypothesis testing |

*Note.* Pre-Registration is available at OSF: https://osf.io/2acqu?view_only=6241cc33bb8949f3b7aa7fc2f8d4f81a

Translation of the Treatment Conditions

*Condition 1 – Human Decision-Making*

**The budget of the European Union**

The annual EU budget is €160 billion (2018). This is a large sum in absolute terms, but represents only 1.02% of the EU's annual economic output.

The money from the EU budget is used in areas where it makes sense to pool resources for the benefit of Europe as a whole, for example:

- the development of transport routes, energy networks and communication links between EU countries,
- the protection of the environment throughout Europe,
- increasing the global competitiveness of the European economy,
- the promotion of transnational groupings of European researchers and scientists.

**Who decides on the use of the funds?**

The decision on the budget for each year is made in two main steps:

1. In a first step, the European Commission prepares a draft budget and submits it to the governments of the member states - represented in the Council of the EU - and to the democratically elected European Parliament.
2. The Commission's budget proposal is then debated, negotiated and, if necessary, adapted in the European Council and the European Parliament. Once the proposal has been accepted by all the institutions involved, the budget for the following year is ready.

EU budget 2018 as pie chart itemized by the financial framework

*Condition 2 – Algorithmic Decision-Making*

**The budget of the European Union**

The annual EU budget is €160 billion (2018). This is a large sum in absolute terms, but represents only 1.02% of the EU's annual economic output.

The money from the EU budget is used in areas where it makes sense to pool resources for the benefit of Europe as a whole, for example:

- the development of transport routes, energy networks and communication links between EU countries,
- the protection of the environment throughout Europe,
- increasing the global competitiveness of the European economy,
- the promotion of transnational groupings of European researchers and scientists.

**Who decides on the use of the funds?**

The decision on the budget for each year is made in two main steps:

1. As a first step, high performance computers of the European Court of Auditors bring together all data available at EU level. Examples are available structural and administrative data from the EU and individual member states, economic and social forecasting models and other data from business and science. On the basis of large data sets, an "Artificial Intelligence" calculates the optimal distribution key of resources for the individual areas of the EU budget within a few hours with the help of so-called machine learning applications.
2. The resulting model is audited by the Court of Auditors and then presented to the President of the European Commission and the Commissioner for Financial Programming and Budget for signature. Thus the budget for the following year is ready.

EU budget 2018 as pie chart itemized by the categories the financial framework

*Condition 3 – Hybrid Decision-Making*

**The budget of the European Union**

The annual EU budget is €160 billion (2018). This is a large sum in absolute terms, but represents only 1.02% of the EU's annual economic output.

The money from the EU budget is used in areas where it makes sense to pool resources for the benefit of Europe as a whole, for example:

- the development of transport routes, energy networks and communication links between EU countries,
- the protection of the environment throughout Europe,
- increasing the global competitiveness of the European economy,
- the promotion of transnational groupings of European researchers and scientists.

**Who decides on the use of the funds?**

The decision on the budget for each year is made in two main steps:

1. In a first step, high performance computers of the European Court of Auditors bring together all data available at EU level. Examples are available structural and administrative data from the EU and individual Member States, economic and social forecasting models and other data from business and science. On the basis of large data sets, an "artificial intelligence" calculates the optimal distribution key of resources for the individual areas of the EU budget within a few hours with the help of so-called machine learning applications.
2. The budget proposal is then debated, negotiated and, if necessary, adapted in the European Commission, the European Council and the European Parliament. Once the proposal has been accepted by all the institutions involved, the budget for the following year is ready.

EU budget 2018 as pie chart itemized by the financial framework.